\documentclass[10pt, notitlepage]{article}
\pdfoutput=1

\usepackage{amssymb}
\usepackage{float}
\usepackage{subfig}
\usepackage{graphicx}
\usepackage[T1]{fontenc}
\usepackage[utf8]{inputenc}
\usepackage{authblk}
\usepackage{amsmath,bm}
\usepackage{multicol}
\usepackage{amsmath}
\usepackage[left=1.2cm,right=1.2cm,top=2.0cm,bottom=2.0cm]{geometry}
\renewcommand\footnotemark{}

\begin{document}

\title{Quantum gravity without vacuum dispersion}
%\title{Vanquishing vacuum dispersion}
\author{D.N. Coumbe}
\affil{\small{\emph{The Niels Bohr Institute, Copenhagen University \\Blegdamsvej 17, DK-2100 Copenhagen Ø, Denmark.\\E-mail: daniel.coumbe@nbi.ku.dk}}} 
\date{\small({\today})}
\maketitle

%%%%%%%%%%%%%%%%%%%%%%%%%%%%%%%%%%%%%%%%%%%%%%%%%%%%%%%%%%%%%%%%%%%%%%%%%%%%%%%%                                                                                                               
\begin{abstract}

A generic prediction of quantum gravity is the vacuum dispersion of light, and hence that a photon’s speed depends on its energy. We present further numerical evidence for a scale dependent speed of light in the causal dynamical triangulation (CDT) approach to quantum gravity. We show that the observed scale dependent speed of light in CDT can be accounted for by a scale dependent transformation of geodesic distance, whose specific functional form implies a discrete equidistant area spectrum. We make two non-trivial tests of the proposed scale transformation: a comparison with the leading-order quantum correction to the gravitational potential and a comparison with the generalised uncertainty principle. In both cases, we obtain the same functional form.

However, contrary to the widespread prediction of vacuum dispersion in quantum gravity, numerous experiments have now definitively ruled out linear vacuum dispersion beyond Planckian energy scales $E_{P}$, and have even constrained quadratic dispersion at the level $\sim 10^{-8}E_{P}$. Motivated by these experimental constraints we seek to reconcile quantum gravity with the absence of vacuum dispersion. We point out that given a scale dependent geodesic distance, a scale dependent time interval becomes essential to maintaining an invariant speed of light. We show how a particular scale dependent time interval allows a photon's speed to remain independent of its energy. 

\vspace{0.5cm}
\noindent \small{PACS numbers: 04.60.Gw, 04.60.Nc}\\
\noindent \small{Keywords: Quantum gravity phenomenology; Lorentz invariance violation; spectral dimension; minimal length scale.}

\vspace{0.5cm}
\end{abstract}

%%%%%%%%%%%%%%%%%%%%%%%%%%%%%%%%%%%%%%%                                                                                                                                                   

\setlength{\columnsep}{25pt}  
\begin{multicols}{2}

\begin{section}{Introduction}

Physics can be characterised as the search for laws that remain invariant under increasingly general transformations. For example, in Newtonian mechanics the laws of physics are invariant under Galilean transformations, in special relativity, they are invariant under the Lorentz transformations, and in general relativity under arbitrary differential coordinate transformations. Each successive group of transformations does not abandon the previous group but rather shows it to be a limiting case of a more general symmetry. For instance, general relativity retains Lorentz invariance as a local symmetry built into the theory at a foundational level via the equivalence principle. In fact, Lorentz symmetry remains the most fundamental symmetry in modern physics, underpinning our two most rigorously tested theories, quantum field theory and general relativity. 

The problem, however, is that combining quantum field theory and general relativity in a straightforward way results in the violation of Lorentz invariance. Mass-energy equivalence combined with the energy-time uncertainty principle implies that the smaller the region of spacetime under consideration the greater the allowed energy of vacuum fluctuations. Hence, a basic prediction of quantum gravity is that as one probes spacetime on ever decreasing distance scales one should observe increasingly large metric fluctuations. Since higher energy, shorter wavelength, photons probe spacetime on shorter time intervals they should encounter proportionately larger metric fluctuations, resulting in an energy dependent speed of light and a deformation or violation of Lorentz invariance. Moreover, quantizing the gravitational field means quantizing spacetime, and therefore presumably quantizing length. Yet the concept of a minimal length seems incompatible with Lorentz invariance: for any quantized minimal length in one observer's rest frame, a second Lorentz boosted observer can always measure a yet shorter length. Maintaining an observer independent minimal length also then appears to require a deformation or violation of Lorentz invariance.

However, after nearly a century of unconstrained theoretical speculation, experiment is now finally able to guide the development of quantum gravity. Recent observations of distant gamma-ray bursts by the \emph{Fermi} space telescope find the speed of light to be independent of energy up to $7.62 E_{P}$ (with a $95 \%CL$) and $\simeq 4.8 E_{P}$ (with a $99 \%CL$)~\cite{Vasileiou:2013vra} for linear dispersion relations, where $E_{P}$ is the Planck energy. The same data even constrains quadratic dispersion relations at the level $\sim 10^{-8}E_{P}$~\cite{Vasileiou:2013vra}. Furthermore, any possible variation in the speed of light is experimentally excluded at the level $\Delta c/c < 6.94\times 10^{-21}$, demonstrating that spacetime remains smooth at energies exceeding the Planck scale~\cite{Nemiroff:2011fk}. A combined analysis of data collected by the \emph{Chandra} and \emph{Fermi} space telescopes in addition to the ground-based \emph{Cherenkov} telescopes have already ruled out a number of approaches to quantum gravity that predict vacuum dispersion~\cite{Vasileiou:2015wja,Diosi:1989hy,Perlman:2014cwa}. An increasingly large number of experiments, using a variety of different techniques, also find results consistent with an energy independent speed of light, including particle decay processes~\cite{Mattingly:2005re}, time-of-flight comparisons~\cite{Antonello:2012be}, neutrino oscillation experiments~\cite{Auerbach:2005tq} and ultra high-energy cosmic ray observations~\cite{Bi:2008yx}.

Furthermore, there exist strong theoretical reasons for preserving an energy independent speed of light. As pointed out by Polchinski~\cite{Polchinski:2011za}, nearly all approaches to quantum gravity containing high-energy vacuum dispersion are already ruled out by precision low-energy tests because of the way such effects scale with energy, with the only possible exception being supersymmetric theories~\cite{Polchinski:2011za}. Vacuum dispersion also permits the existence of a preferred observer, which would violate one of the oldest and most reliable physical principles, the relativity principle. Moreover, the existence of vacuum fluctuations imply a host of implausible scenarios such as causality violations~\cite{Mattingly:2005re} and closed time-like curves~\cite{Mattingly:2005re}. %Abandoning a constant speed of light should be an absolute last resort. 

Individually, no experimental result or theoretical argument can ever definitively rule out vacuum dispersion. However, the accumulative weight of evidence is now strongly suggestive: experiment is unambiguously telling us that spacetime remains a smooth manifold on much shorter distances than many approaches to quantum gravity predict. How then might we reconcile the near ubiquitous prediction of vacuum dispersion in quantum gravity with the growing body of contrary experimental evidence? This work aims to answer this question by exploring a particularly simple way of eliminating vacuum dispersion in quantum gravity.

\end{section}

%%%%%%%%%%%%%%%%%%%%%%%%%%%%%%%%%%%%%%%%%%%%%%%

\begin{section}{Characterising vacuum dispersion}\label{vti}

Although there is currently no precise mathematical description of the predicted extent of vacuum dispersion, a useful characterisation of the expected deformed dispersion relation can be obtained by assuming the deformation admits a series expansion at small energies $E$ relative to the energy scale $E_{QG}$ at which quantum gravitational effects become significant (presumably on the order of $E_{P}$)~\cite{Vasileiou:2013vra,AmelinoCamelia:1999zc}. Such a series expansion yields a deformed dispersion relation

\begin{equation}
E^{2} \simeq p^{2} \left(1 - \sum_{n=1}^{\infty} \pm \left(\frac{E}{E_{QG}}\right)^{n} \right),
\label{DeformedDispQG}
\end{equation}

\noindent which implies an energy dependent speed of light

\begin{equation}
c(E)\simeq \sqrt{1 - \sum_{n=1}^{\infty} \pm \left(\frac{E}{E_{QG}}\right)^{n}},
\label{DeformedDispQG}
\end{equation}

\noindent where the conventional speed of light in the zero-energy limit $c$ is set equal to unity (hereafter we set $\hbar=c=1$, unless stated otherwise). The $\pm$ ambiguity in Eq.~(\ref{DeformedDispQG}) denotes either subluminal $(+1)$ or superluminal $(-1)$ motion~\cite{Vasileiou:2013vra}. In effective field theory $n=d-4$, where $d$ is the dimension of the leading order operator. The $n=1$ term then comes from a dimension 5 operator~\cite{Myers:2003fd}. Since odd values of $d$ have been shown to violate CPT invariance the leading order term is expected to be $n=2$~\cite{Colladay:1996iz,Colladay:1998fq}.\interfootnotelinepenalty=10000 \footnote{\scriptsize Although it is currently unclear how reliable effective field theory will prove to be in the description of Planckian scale physics, and hence whether it is valid to \emph{a priori} restrict the symmetries of quantum gravity in this way.}

In subsection~\ref{dimred} we aim to more precisely characterise the vacuum dispersion of light by studying a specific approach to quantum gravity known as causal dynamical triangulations (CDT), and in subsection~\ref{area} we show that the resulting functional form suggests the discretisation of area and explore how this may be used to characterise vacuum dispersion in a model-independent setting. 

%%%%%%%%%%%%%%%%%%%%%%%%%

\begin{subsection}{Dimensional reduction and vacuum dispersion}\label{dimred}

Assuming only established principles of general relativity and quantum field theory, and including few additional ingredients, causal dynamical triangulations (CDT) define a particularly simple approach to quantum gravity (see~\cite{Ambjorn05} for more details). The simplicity of CDT is also its greatest strength; numerical experiments in CDT are capable of providing reliable and robust insights into Planck scale physics, free from the potential pit-falls of assuming more exotic ingredients.%\interfootnotelinepenalty=10000 \footnote{\scriptsize This subsection closely follows Ref.~\cite{Coumbe:2015zqa} and significantly improves on the numerical results presented therein.}

One striking insight to come from these numerical experiments suggests the dimension of spacetime dynamically reduces from $\sim$4 on macroscopic scales to $\sim$2 on microscopic scales~\cite{Ambjorn:2005db,Coumbe:2014noa}. This observation has sparked considerable interest in dimensional reduction throughout the quantum gravity community, with exact renormalisation group approaches~\cite{Lauscher:2005qz}, Hořava-Lifshitz gravity~\cite{Horava:2009if}, noncommutative geometry~\cite{Arzano:2014jfa,Benedetti:2008gu}, loop quantum gravity~\cite{Modesto:2008jz} and string theory~\cite{Atick:1988si,Calcagni:2013eua} all reporting a similar reduction in the number of spacetime dimensions near the Planck scale. %Dimensional reduction is now a ubiquitous feature of quantum gravity.   

Evidence for dimensional reduction has come mainly from calculations of the spectral dimension, a measure of the effective dimension of a manifold over varying length scales. The spectral dimension $D_{S}$ is related to the probability $P_{r}\left(\sigma\right)$ that a random walk will return to the origin after $\sigma$ diffusion steps, and is defined by\interfootnotelinepenalty=10000 \footnote{\scriptsize Equation~(\ref{specD}) is strictly only valid for infinitely flat Euclidean space. However, one can still use this definition of the spectral dimension to calculate the fractal dimension of a curved space, or finite volume, by factoring in the appropriate corrections (see Refs.\cite{Ambjorn:2005db,Coumbe:2014noa} for more details).}

\begin{equation}\label{specD}
D_{S}=-2\frac{d\rm{log} P_{r}\left(\sigma\right)}{d\rm{log}\sigma}.
\end{equation}

\noindent In momentum space $P_{r}\left(\sigma\right)$ is given by

\begin{equation}\label{retMom}
P_{r}\left(\sigma\right)=\int e^{\sigma \bigtriangleup_{p} } d\mu,
\end{equation}

\noindent where $d\mu=dE d^{3}\vec{p} / (2\pi)^{4}$ and $\bigtriangleup_{p}$ are the invariant measure and Laplace operator in momentum space, respectively~\cite{Mielczarek:2015cja}. An undeformed dispersion relation in 4-dimensional Euclidean space gives $\bigtriangleup_{p}=-E^{2}-p^{2}=0$, which via Eqs.~(\ref{specD}) and~(\ref{retMom}) maintains a scale invariant spectral dimension of 4. Crucially, however, if the spectral dimension $D_{S}$ varies as a function of $\sigma$ then $\bigtriangleup_{p} \neq 0$, and we must have a deformed dispersion relation~\cite{Mielczarek:2015cja}. Therefore, a reduction of the spectral dimension implies either a violation or deformation of Lorentz invariance. 

One can also start from specific modified dispersion relations and show that they can lead to a reduction of the spectral dimension~\cite{Sotiriou:2011aa}. It can be shown that the reduction of the spectral dimension reported in a large variety of approaches to quantum gravity can be derived from a modified dispersion relation~\cite{Amelino-Camelia:2013tla,Sotiriou:2011aa,Horava:2009if} 

\begin{equation}\label{gammadisp}
E^{2}=p^{2}\left(1+(\lambda p)^{2\gamma}\right),
\end{equation}

\noindent which implies a variable speed of light

\begin{equation}\label{gammaspeed}
c(\lambda p)= \sqrt{1+(\lambda p)^{2\gamma}},
\end{equation}

\noindent where $\lambda p$ is a momentum scale and $\gamma$ a positive integer. 

In a spacetime with $(d+1)$ Hausdorff dimensions the deformed dispersion relation of Eq.~(\ref{gammadisp}) has been shown~\cite{Sotiriou:2011aa,Horava:2009if} to lead to a short distance ultraviolet (UV) spectral dimension of

\begin{equation}\label{gammauv}
D_{S}^{UV}=1+\frac{d}{1+\gamma},
\end{equation}

\noindent which gives $D_{S}^{UV}=2$ when $\gamma=2$. Therefore, a scale dependent spectral dimension implies a deformed dispersion relation, and conversely certain deformed dispersion relations imply a scale dependent spectral dimension. We now focus on the specific case of CDT quantum gravity, revisiting and extending the work of Ref.~\cite{Coumbe:2015zqa} by providing additional numerical evidence for a scale dependent speed of light associated with a reduction of the spectral dimension.

The most studied point in the parameter space of CDT has been shown to have a scale dependent spectral dimension that appears to be most accurately described by the functional form

\begin{equation}
D_{S}=a-\frac{b}{c+\sigma},
\label{funcform}
\end{equation}

\noindent where $a$, $b$ and $c$ are free fit parameters \cite{Ambjorn:2005db,Coumbe:2014noa}. This particular functional form of the spectral dimension is also supported by purely analytic methods~\cite{Giasemidis:2012qk,Giasemidis:2012rf}. Integrating Eq.~(\ref{funcform}) gives a return probability 

\begin{equation}
P_{r}\left(\sigma\right)=\frac{1}{\sigma^{a/2}\left(1+\frac{c}{\sigma}\right)^{\frac{b}{2c}}}.
\end{equation}

\noindent CDT simulations yield a fit to the data with $a=4.02$, $b=119$ and $c=54$~\cite{Ambjorn:2005db}. A more recent study at the same point in the CDT parameter space also gives similar results, namely $a=4.06$, $b=135$ and $c=67$ \cite{Coumbe:2014noa}. Both of these independent calculations of the spectral dimension therefore find that $a\simeq 4$ and $b/2c \simeq 1$, and so

\begin{equation}
P_{r}\left(\sigma\right) \simeq \frac{1}{\sigma^{2}+c \sigma}.
\label{RetProb2}
\end{equation}

The probability of return for infinitely flat 4-dimensional Euclidean space with no dimensional reduction is $P\left(\sigma\right)=\sigma^{-2}$. Since the path length of a diffusing particle is proportional to the number of diffusion steps $\sigma$, we ask what function $\Gamma_{+}(\sigma)$ rescales the path length such that we obtain the probability of return found in CDT, namely that of Eq.~(\ref{RetProb2})? To answer this question we form the equation

\begin{equation}
\frac{1}{\Gamma_{+}\left(\sigma\right)^{2}\sigma^{2}}=\frac{1}{\sigma^{2}+c \sigma},
\end{equation}

\noindent finding that

\begin{equation}
\Gamma_{+}\left(\sigma\right)=\sqrt{1+\frac{c}{\sigma}}.
\label{GammaFuncSig}
\end{equation}

\noindent Therefore, the appearance of dimensional reduction can be attributed to a path length $l$ that grows as a function of decreasing $\sigma$ according to $l \rightarrow l \Gamma_{+}(\sigma)$. In fact, such a scale dependent path length is a generic feature of fractal curves \cite{AbbottWise}. 

In flat space, $\sigma$ corresponds to probing spacetime at a linear scale $r=\sqrt{\sigma}$, where a large $\sigma$ value conforms to a large linear distance from the origin of the diffusion process, and a small $\sigma$ to a short distance~\cite{Ambjorn05}. As alluded to previously, a spectral dimension that varies with the linear distance scale $r$ implies either a systematic violation, a non-systematic violation or a deformation of Lorentz invariance~\cite{Mattingly:2005re,Coumbe:2015zqa}. Thus, identifying $\sigma$ with $r^{2}$ comes with the radical implication that Lorentz invariance is at the very least deformed, a point the reader should be aware of. In section~\ref{varyingtint} we present a method for reconciling a scale dependent spectral dimension with Lorentz invariance. 

Assuming the free fit parameter $c$ in Eq.~(\ref{GammaFuncSig}) can be expressed in Planck units by $c=Al_{p}^{2}$ as suggested in Ref. \cite{Ambjorn:2005db}, where $l_{p}$ is the Planck length and $A$ is a numerical constant of order unity, and that $\sigma=r^{2}$ then Eq.~(\ref{GammaFuncSig}) becomes

\begin{equation}\label{Gammar}
\Gamma_{+}(r)=\sqrt{1+ \frac{A l_{p}^{2}}{r^2}}.
\end{equation}

\noindent Thus, as we resolve spacetime on ever decreasing radial scales $r$ the path length $l$ of the massless diffusing particle increases according to $l \rightarrow l \Gamma_{+}(r)$. For a geodesic path, such as that of light, this is equivalent to radial distance $r$ scaling according to $r \rightarrow r \Gamma_{+}(r) \equiv r'$. Given a scale invariant time interval, such a variable path length results in a scale dependent speed of light

\begin{equation}\label{modsol}
c_{m}\left(r\right)= \Gamma_{+}\left(r\right).
\end{equation}

To support this picture, one can even explicitly track the path a fictitious diffusing particle traces out in a given ensemble of CDTs. The diffusing test particle begins in a randomly chosen simplex and diffuses throughout the geometry by making $\sigma$ jumps between adjacent simplices. By mapping the random walk of the diffusing particle within the ensemble we can extract information about how the path length varies as a function of geodesic distance from its origin. The particle's path length is equal to the number of diffusion steps $\sigma$ multiplied by the average distance between adjacent simplices, which we encode in the constant of proportionality $\zeta$. CDT distinguishes between space-like and time-like links on the lattice such that one can define an explicit foliation of the lattice into space-like hypersurfaces. CDT thereby introduces a time coordinate within the ensemble via space-like hypersurfaces at fixed time intervals $t=0$, $t=1$, ..., $t=N$, defining a causal slice of spacetime of duration $t=N$. The elapsed time for a particle to diffuse between two arbitrary points is then the number of times it intersects a space-like hypersurface, which we call $t_{d}$. Thus, we define an effective velocity $v_{d}$ for the diffusing particle via the ratio\interfootnotelinepenalty=10000 \footnote{\scriptsize Since the spectral dimension is defined in Euclidean signature a speed of light cannot be defined and so the diffusion rate $v_{d}$ is an effective velocity.}

\begin{equation}
v_{d}=\frac{\zeta \sigma}{t_{d}}.
\end{equation}

Figure~\ref{SOLdiff} shows $v_{d}$ averaged over $10^{6}$ different diffusion processes for the canonical point in the de Sitter phase of CDT using two different lattice volumes, and for a constant of proportionality $\zeta=0.18$. Figure~\ref{SOLdiff} supports and improves the findings of Ref.~\cite{Coumbe:2015zqa}. The important feature of Fig.~\ref{SOLdiff} is that the measured velocity of the diffusing particle $v_{d}$ in a typical CDT ensemble of triangulations is superluminal and closely matches the scale dependent speed of light $c_{m}(\sigma)=\Gamma_{+}(\sigma)$ predicted by dimensional reduction.

%,natwidth=610,natheight=642
\begin{figure}[H]
\centering
\includegraphics[width=1.0\linewidth]{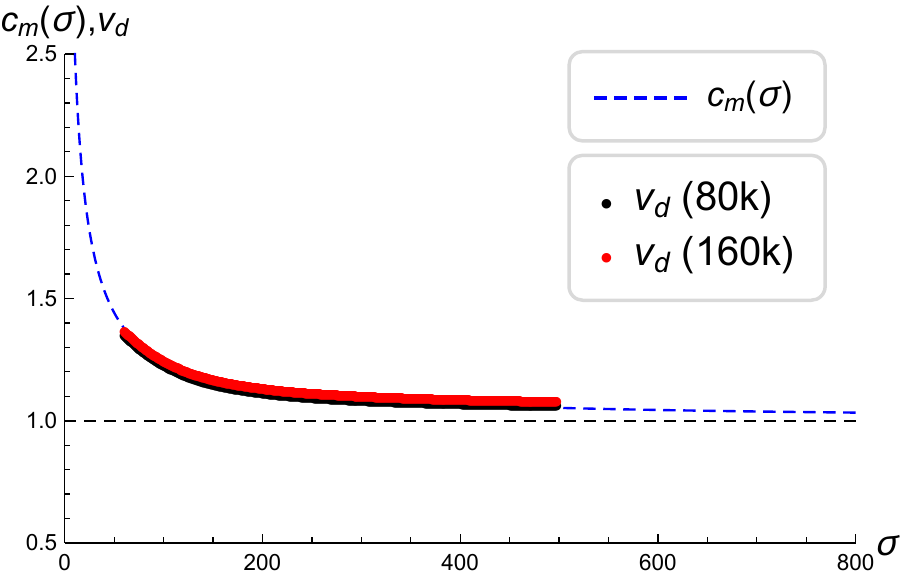}
\caption{\small A comparison between the modified speed of light $c_{m}(\sigma)=\Gamma_{+}(\sigma)$ with $c=54$ as determined from the study of dimensional reduction in Ref.~\cite{Ambjorn:2005db} (the dashed blue curve) and numerical measurements of the effective velocity $v_{d}$ determined by averaging over $10^{6}$ diffusion processes using 80,000 (black data points) and 160,000 (red data points) simplices with $\zeta=0.18$. The dashed black line denotes the conventional speed of light.}
\label{SOLdiff}
\end{figure}

Probing spacetime on smaller distance scales requires greater energies. Probing spacetime at a scale $r$ requires an energy $E = 2\pi / \lambda = 2\pi / (Br)$, where $B$ is a constant of proportionality of order unity. The function $\Gamma_{+}(r)$ given by Eq.~(\ref{Gammar}) can then be written in terms of energy as

\begin{equation}
\Gamma_{+}(E)=\sqrt{1+\frac{A B^{2}}{4\pi^{2}}\frac{E^{2}}{E_{P}^{2}}},
\end{equation}

\noindent which via Eq.~(\ref{modsol}) implies an energy dependent speed of light 

\begin{equation}\label{dimredsol}
c(E)=\sqrt{1+\frac{A B^{2}}{4\pi^{2}}\left(\frac{E}{E_{P}}\right)^{2}}.
\end{equation}

\noindent Equation~(\ref{dimredsol}) is strikingly similar to the putative energy dependent speed of light given by Eq.~(\ref{DeformedDispQG}), and conforms with the expectation for the leading order term based on CPT invariance.

%%%%%%%%%%%%%%%%%%%%%%%%%%%%%%%%%%%%%%%%%%%%%%%%%%%%%%%%%%%%%%%%%%%%%

\end{subsection}

\begin{subsection}{Area discretisation and vacuum dispersion}\label{area}

%In subsection~\ref{dimred} we argued that the reduction of the spectral dimension observed in CDT can be explained by a path length $l$ that transforms as a function of the scale $r$ with which we probe spacetime according to $l \rightarrow l \Gamma_{+}(r)\equiv l'$. But why this particular scaling of path length? We seek an answer to this question by applying the scaling relation inferred from CDT experiments to radial distance $r$. In this scenario, $r$ scales according to $r'=r \Gamma_{+}(r)$, from which it follows that

In subsection~\ref{dimred} we argued that the reduction of the spectral dimension observed in CDT can be explained by a path length $l$ that transforms as a function of the scale $r$ with\
 which spacetime is probed according to $l \rightarrow l \Gamma_{+}(r)\equiv l'$. But why this particular scaling of path length? In this subsection we explore a possible answer to this question.% by applying the transformation inferred from the diffusion of massless particles in CDT to the simpler case of photon propagation.

We first consider applying the transformation inferred from the diffusion of massless particles in CDT to the simpler case of photon propagation. Since light follows geodesic paths in spacetime, the transformation of path length implied by the CDT data is then equivalent to scaling geodesic distance according to $r \rightarrow r \Gamma_{+}(r)\equiv r'$, which is remarkably similar to the transformation found in Refs.~\cite{Kothawala:2013maa,Padmanabhan:2015vma}. Using this scale transformation and the definition of $\Gamma_{+}(r)$ it can be shown that

\begin{equation}
r'^{2} - r^{2} =A l_{P}^{2}.
\end{equation}

\noindent Similarly, we can define a new radius $r''$ by applying the same scale transformation to $r'$, namely $r''=r' \Gamma_{+}(r')$, obtaining

\begin{equation}
r''^{2} - r'^{2} =A l_{P}^{2}.
\end{equation}

\noindent Thus, the scale transformation $r \rightarrow r \Gamma_{+}(r)\equiv r'$ defines an equidistant area spectrum, where the surface area differs by an integer multiple of the Planck area.

%%%%%%%%%%%%%%%%%%
%%%%%%%%%%%%%%%%%%

The discretisation of area appears consistently across a wide range of physical scenarios. Discretising the area of lightlike surfaces was first proposed in Bekenstein's seminal work on black holes~\cite{Bekenstein:1974ab,Bekenstein:1973ur,Bekenstein:1974ax}. Bekenstein found that the event horizon of a non-extremal black hole behaves like a classical adiabatic invariant, which via Ehrenfest's theorem~\cite{Born69ab} corresponds to a discrete eigenvalue spectrum~\cite{Bekenstein:1972bj,Bekenstein:1973ur}. We can further explore the possible connection between the proposed scaling of geodesic distance and a discrete equidistant area spectrum by considering a physical scenario similar to that proposed by Bekenstein, but in the more general setting of flat spacetime. Before doing so we briefly recap Bekenstein's argument:

Consider a single photon of energy $E=\omega$ falling into a black hole of Schwarzschild radius $r_{S}=2Gm$. For simplicity we wish to add only one bit of information to the black hole, and so we must make the wavelength of the photon sufficiently large such that its entry point is delocalized over the entire horizon. In this way we can have no information on where exactly the photon entered the horizon, only whether it did. We therefore consider a photon of wavelength $\lambda = B r_{S}$, where again $B$ is a constant of proportionality of order unity. The energy of this photon is $E=2 \pi /(B r_{S})$, which increases the mass of the black hole by $\delta m=2\pi / (B r_{S})$. This change of mass translates into a change in horizon area

\begin{equation}\label{swarzs}
\delta A_{S}=r_{S}\delta r_{S}=\frac{4\pi}{B}  l_{P}^{2},
\end{equation}

\noindent where we have used $l_{P}^{2}=G$.

Therefore, the surface area of an event horizon increases by a constant factor multiplied by the Planck area every time we add one bit of information. If we were to sequentially add $n$ bits of information the allowed values of horizon area would then define a discrete equidistant area spectrum

\begin{equation}
A_{n}=\frac{4\pi}{B} nl_{P}^{2},\quad  n \in \mathbb{N}.
\label{areaspectrum}
\end{equation}

\noindent Now, if in flat spacetime the classical notion of an exact geodesic distance $r$ between two spacetime points $x_{1}$ and $x_{2}$ is to be replaced by a modified distance

\begin{equation}\label{geo1}
r'^{2}(x_{1},x_{2})= r^{2}(x_{1},x_{2}) \left[ 1 + \frac{\zeta l_{P}^{2}}{\lambda^{2}}\right ],
\end{equation}

\noindent where $\zeta$ is a constant of proportionality, $l_{P}$ is the Planck length and $\lambda$ is the resolution of the measurement, then we can consider an argument similar to Bekenstein's but in a more general setting.

\noindent Let $x_{1}$ and $x_{2}$ be any two points in flat spacetime. Let $x_{1}$ define the origin of a spherical region of spacetime and let $x_{2}$ be a point on its surface. Just as in Bekenstein's argument we wish to add only one bit of information to the spherical region, and so we must make the wavelength of the photon sufficiently large such that its entry point is delocalized over the entire region. This prevents us from obtaining information on where exactly the photon entered the region, only whether or not it did. We therefore use light of wavelength $\lambda = B r(x_{1},x_{2})$ to measure the distance between the two points, where $B$ is a constant of proportionality. (Alternatively, light of wavelength $\lambda \approx r(x_{1},x_{2})$ defines the minimal resolution at which the two points $x_{1}$ and $x_{2}$ can be distinguished as individuals, and hence making a measurement with a resolution $\lambda \approx r(x_{1},x_{2})$ encodes one bit of binary information.)

\noindent Equation~(\ref{geo1}) then tells us that probing a classical distance of $r(x_{1},x_{2})$ with resolution $\lambda = B r(x_{1},x_{2})$ gives a modified distance

\begin{equation}
r'^{2}(x_{1},x_{2}) = r^{2}(x_{1},x_{2}) \left[ 1 + \frac{\zeta l_{P}^{2}}{r^{2}(x_{1},x_{2})}\right ].
\end{equation}

\noindent Thus, the measurement process perturbs the original distance $r(x_{1},x_{2})$ such that points $x_{1}$ and $x_{2}$ now define a larger sphere of radius $r'(x_{1},x_{2})$, where the surface area of the resulting sphere is greater than the original by one Planck area. If this scenario is correct then adding one bit of information to a spherical region of flat spacetime increases the surface area of the bounding surface by exactly one Planck area. If we were to sequentially add $n$ bits of information the allowed values of the spherical region would then define the same discrete equidistant area spectrum of Eq.~(\ref{areaspectrum}) found by Bekenstein. 

%%%%%%%%%%%%%%%%%%%%%%%%
%%%%%%%%%%%%%%%%%%%%%%%%

%The restriction to a fixed foliation in CDT defines a fixed set of equally spaced time intervals. The CDT data indicates that the massless diffusing test particles traverse unequal distances in equal times, giving rise to the scale dependent speed shown in Fig.\ref{}. Lifting the restritction of a fixed foliation allows us to apply the scaling relation inferred from the CDT experiments to the propagation of massless photons along geodesic paths.  

%If we drop the restritction of a fixed foliation and instead consider the propagation of photons subject to the same scaling found in the CDT experiments then the photons would cover unequal geodesic distances in equal times.

%, which for a geodesic path is equivalent to radial distance $r$ scaling according to $r \rightarrow r \Gamma_{+}(r) \equiv r'$. But why this particular scaling of geodesic distance? As we shall show, it turns out that this is precisely the scaling relation one obtains if area is dicretised in multiples of the Planck area.

%By integrating the expression $dr'=\Gamma_{+}\left(r\right)dr$ we obtain $r=r'\Gamma_{+}\left(r'\right)$, which defines a discrete spectrum of radii.

%The well-studied properties of event horizons also seem applicable in a much more general context, with de Sitter and Rindler horizons sharing many of the salient features~\cite{Jacobson:2003wv}. In addition to horizons, 

Motivation for the discretisation of area can also be derived from the robust prediction of a zero-point length in quantum gravity~\cite{Padmanabhan:1996ap,Padmanabhan:2015vma}, from fluctuations of the conformal factor~\cite{padmanabhan1985physical}, from holographic considerations~\cite{'tHooft:1994un,Bousso:2002ju}, from entanglement entropy~\cite{Eisert:2008ur}, from loop quantum gravity~\cite{Khriplovich:2004fd,Rovelli:1994ge}, and from string-inspired approaches~\cite{Visser:2012zi,Halyo:1996xe}. Primarily motivated by the scaling relation inferred from the CDT experiments, but also by the wide-ranging evidence for the discretisation of area, we now investigate the propagation of lightlike surfaces discretised in multiples of the Planck area as a way of further characterising the form of vacuum dispersion.

%Motivated by the specific functional form of geodesic distance scaling inferred from our numerical CDT experiments  

%In this work we are motivated to investigate 

%, finding that one obtains the same transformation of geodesic distance necessary to explain dimensional reduction in CDT. We wish to make it clear to the reader that although the discretisation of area appears to be a consistent feature of quantum gravity there is no precident for specifically discretising the area of lightlike surfaces.%, and we explore this possibility primarily because it is suggested by the CDT data. 

%An enormous amount of evidence supporting the equidistant area spectrum of black hole event horizons has accumulated over the last several decades~\cite{Bekenstein:1973ur,Bekenstein:1974ab,Gour:2002pj}. 

%By considering the propagation of light from an arbitrary point we now show that one obtains an identical transformation of geodesic distance if the surface area of its spherical wavefront is discretised in multiples of the Planck area.

Let a spherical light wave propagate radially outwards from a given spacetime point $P$ and have a discretised surface area

\begin{equation}
A_{n}= \frac{4\pi}{B} n l_{P}^{2},\quad  n \in \mathbb{N},
\label{areaspectrum2}
\end{equation}

\noindent where $B$ is a numerical constant of order unity. Consider a coordinate frame whose origin coincides with the point P, in which an observer probes the spherical light wave at a geodesic distance $r_{n}$, as shown schematically in Fig.~\ref{LightWave}. Applying Eq.~(\ref{areaspectrum2}) to a spherical light wave of area $A_{n}=4\pi r_{n}^{2}$ gives

\begin{equation}
r_{n}=\sqrt{\frac{n}{B}}l_{P}.
\label{rn}
\end{equation}

The area discretisation enforced by Eq.~(\ref{areaspectrum2}) therefore means that the radius of the spherically expanding lightlike surface must be a scale dependent discrete variable. The next largest radial coordinate permitted by the constraint of an equidistant area spectrum is

\begin{equation}
r_{n+1}=\sqrt{\frac{n+1}{B}}l_{P}.
\label{rn}
\end{equation} 

\noindent The smallest factors by which the radius can change at a scale $r_{n}$ are then\interfootnotelinepenalty=10000 \footnote{\scriptsize Note that one is not only restricted to considering neighbouring steps on the discrete area scale. In general one obtains the ratio $\frac{r_{n+j}}{r_{n}}=\sqrt{1+\frac{j}{B}\frac{l_{P}^{2}}{ r_{n}^{2}}}$, where $j$ can be any positive integer.}

\begin{equation}
\frac{r_{n+1}}{r_{n}}=\sqrt{1+\frac{1}{B} \frac{l_{P}^{2}}{r_{n}^{2}}} \equiv \Gamma_{+}\left(r_{n}\right)
\label{Gamma2}
\end{equation}

\noindent and 

\begin{equation}
\frac{r_{n}}{r_{n-1}}=\frac{1}{\sqrt{1-\frac{1}{B} \frac{l_{P}^{2}}{r_{n}^{2}}}} \equiv \Gamma_{-}\left(r_{n}\right).
\label{Gamma3}
\end{equation}

In this scenario, the spherically expanding light wave propagates by making a sequence of jumps between the allowed spectrum of surface areas, i.e. it first traverses the distance $\delta r_{n}\equiv r_{n}-r_{n-1}$ then $\delta r_{n+1}\equiv r_{n+1}-r_{n}$, etc. (see Fig.~\ref{LightWave}). Using Eqs.~(\ref{Gamma2}) and~(\ref{Gamma3}) we obtain the finite differences

\begin{equation}
\delta r_{n} \equiv r_{n}-r_{n-1}=r_{n} \left(1-\frac{1}{\Gamma_{-}\left(r_{n}\right)}\right)
\label{finitediff1}
\end{equation}

\noindent and

\begin{equation}
\delta r_{n+1} \equiv r_{n+1}-r_{n} = r_{n}\left( \Gamma_{+}\left(r_{n}\right) -1 \right).
\label{finitediff2}
\end{equation}

\noindent The propagation speed over the distance $\delta r_{n}$ is then $c(r_{n})=\delta r_{n} / \delta t_{n}$ and over the distance $\delta r_{n+1}$ it is $c(r_{n+1})=\delta r_{n+1}/ \delta t_{n+1}$, where $\delta t_{n}$ and $\delta t_{n+1}$ are the time it takes to traverse the distance $\delta r_{n}$ and $\delta r_{n+1}$, respectively.

\begin{figure}[H]
\centering
\includegraphics[width=0.7\linewidth]{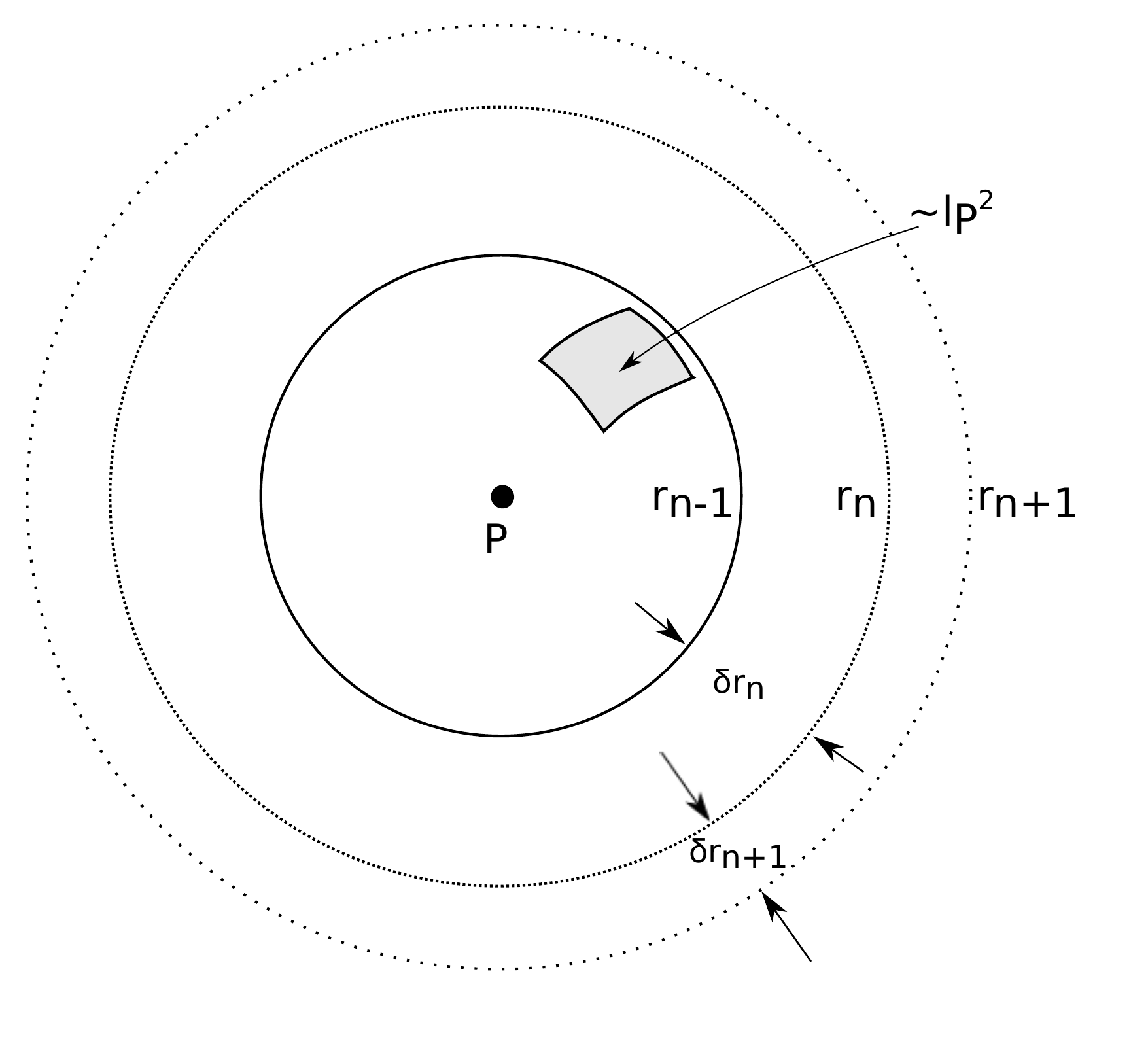}
\caption{\small A schematic representation of a spherically expanding light wave with an equidistant area spectrum.}
\label{LightWave}
\end{figure}

It is expected that an observer $O$ should measure the same time interval between two events originating from the same point $P$, regardless of the geodesic distance of the observer from $P$, assuming asymptotically flat spacetime and that $O$ and $P$ have zero relative motion. For this expectation to be realised the interval between sequential waves measured at two different distances from $P$ must be equal, i.e. $\delta t_{n}=\delta t_{n+1}$. In this subsection we therefore assume that each jump between discrete radial coordinates takes the same amount of time, i.e. $\delta t_{n}=\delta t_{n+1}$, as per our current physical expectations. Assuming fixed time intervals also aids the comparison with diffusion in CDT, where the lattice is foliated into time slices of fixed duration. We remove this constraint in subsections~\ref{DR2} and~\ref{Area2}. 

Under the assumption $\delta t_{n}=\delta t_{n+1}$ the ratio of propagation speeds is

\begin{equation}                                                                                                                                                                        
\frac{c(r_{n})}{c(r_{n+1})}=\frac{\delta r_{n}}{\delta r_{n+1}}= \frac{ 1-\frac{1}{\Gamma_{-}\left(r_{n}\right) }}   {\Gamma_{+}\left(r_{n}\right) -1}.
\label{RfiniteDilation}                                                                                                                                                                 
\end{equation}         

\noindent Differentiating $r_{n}$ with respect to $r_{n+1}$ gives the infinitesimal version of Eq.~(\ref{RfiniteDilation}) as

\begin{equation}
\frac{c(r_{n})}{c(r_{n+1})}=\frac{d r_{n}}{d r_{n+1}}=\Gamma_{+}\left(r_{n}\right),
\label{diffr}
\end{equation}

\noindent where we have again assumed the propagation times are equal, i.e. $dt_{n}=dt_{n+1}$. The infinitesimal ratio of Eq.~(\ref{diffr}) is a valid approximation of the finite ratio given by Eq.~(\ref{RfiniteDilation}) for $r \gtrsim l_{P}$, as can be seen in Fig.~\ref{RatioComp}. In the large distance limit $c(r_{n+1})$ will asymptotically approach the conventional speed of light $c=1$, and since probing spacetime at a scale $r_{n}$ requires an energy $E_{n}=2\pi /(Br_{n})$ we obtain an energy dependent speed of light

\begin{equation}
c(E_{n}) =\sqrt{1+ \frac{B}{4\pi^{2}} \frac{E_{n}^{2}}{E_{P}^{2}}} \equiv \Gamma_{+}\left(E_{n}\right).
\label{diffE}
\end{equation}

\begin{figure}[H]                                        
\centering                                                                                                                                                                              
\includegraphics[width=1.0\linewidth]{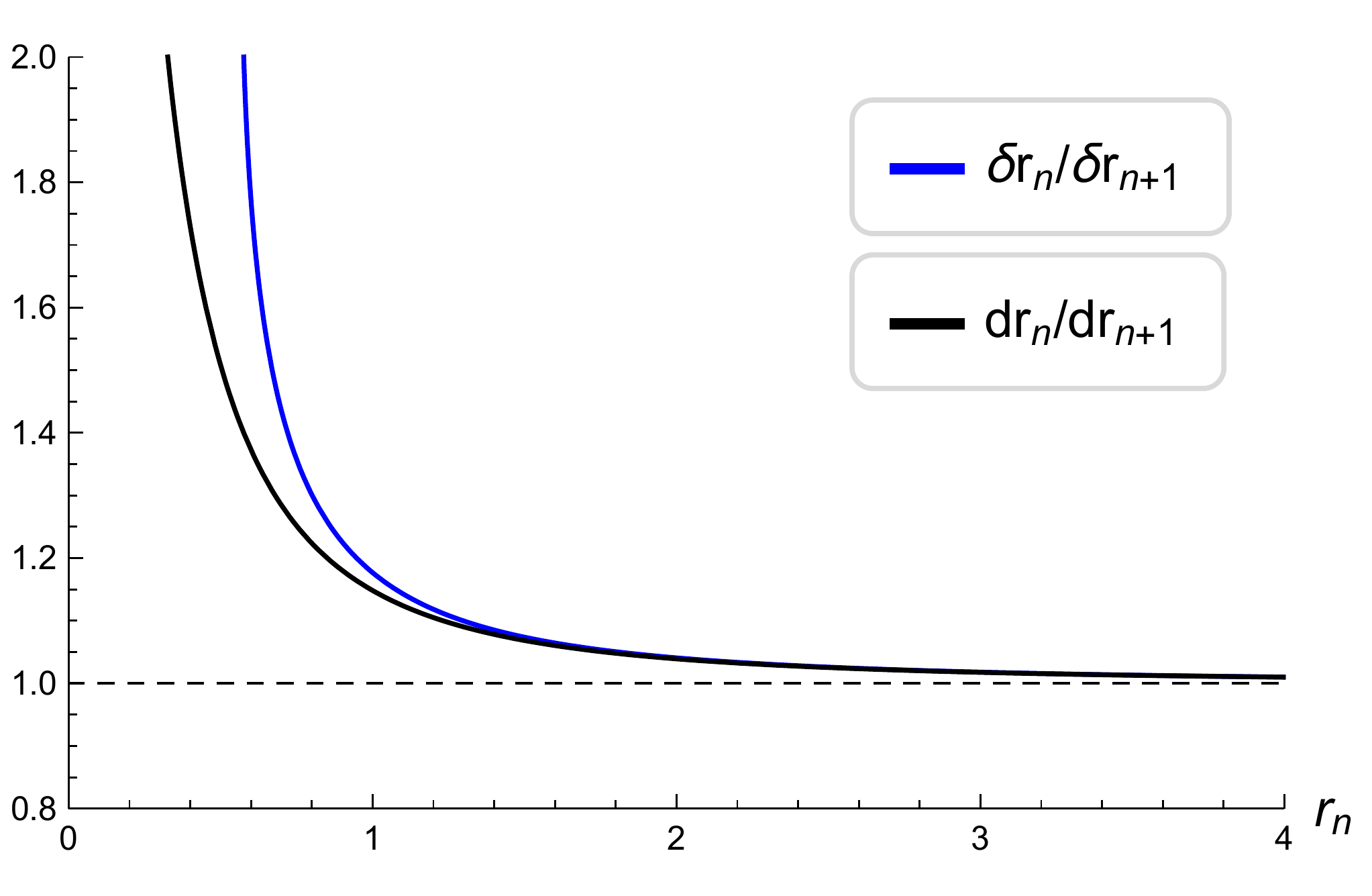}                                                                                                                             
\caption{\small A comparison of $\delta r_{n} / \delta r_{n+1}$ and $dr_{n} / dr_{n+1}$ as a function of scale $r_{n}$ (in units of $l_{P}$).}
\label{RatioComp}                                                                                                                                                                      
\end{figure}    

%Interestingly, the scale dependent speed of light described by Eq.~(\ref{diffr}) is identical to the functional form of Eq.~(\ref{modsol}) associated with dimensional reduction in lattice quantum gravity~\cite{Coumbe:2015zqa}, modulo a numerical constant, suggesting an underlying connection between the discretisation of area and the appearance of dimensional reduction. 

The fact that the measured effective velocity of diffusing particles in CDT closely matches Eq.~(\ref{diffr}) may be interpreted as providing additional evidence for the discretisation of surface area. The near ubiquitous appearance of dimensional reduction in quantum gravity may then be a consequence of the discretisation of area coupled with the implicit assumption that time is scale invariant, as we have assumed in this subsection. In section~\ref{varyingtint} we remove the assumption that time must be scale invariant and compute the consequences.  

Before proceeding we wish to briefly highlight a particular property of the scale transformation described in this subsection. Using the identity

\begin{equation}                                                                                                                                                                        \
\sqrt{1+\frac{1}{n}} \equiv \frac{1}{\sqrt{1-\frac{1}{n+1}}},
\end{equation}                                                                                                                                                                          \

\noindent we can express the scaling function $\Gamma_{+}\left(r_{n}\right)$ in terms of $r_{n+1}$, namely

\begin{equation}
\Gamma_{+}\left(r_{n}\right) = \frac{1}{\sqrt{1-\frac{1}{B}\frac{ l_{P}^{2}}{r_{n+1}^{2}}}} \equiv \Gamma_{-}\left(r_{n+1}\right).
\label{identity}
\end{equation}

\noindent The function $\Gamma_{-}\left(r_{n+1}\right)$ is strikingly similar to the function $\gamma=\left(1-\left(v^{2}/c^{2}\right)\right)^{-1/2}$ associated with special relativity. Specifically, $l_{P}/ \sqrt{B}$ defines the smallest possible scale in $\Gamma_{-}\left(r_{n+1}\right)$, just as $c$ defines the largest possible velocity in $\gamma$. Together, $\gamma$ and $\Gamma_{-}\left(r_{n+1}\right)$ then imply a maximal speed $c$ and a minimal length scale $l_{P}/ \sqrt{B}$, as per the aims of doubly special relativity~\cite{AmelinoCamelia:2002wr}.

\end{subsection}

\end{section}

%%%%%%%%%%%%%%%%%%%%%%%%%%%%%%%%%%%%%%%%%%%%%%%                                                                                                                                         
%%%%%%%%%%%%%%%%%%%%%%%%%%%%%%%%%%%%%%%%%%%%%%%                                                                                                                                         
%%%%%%%%%%%%%%%%%%%%%%%%%%%%%%%%%%%%%%%%%%%%%%%                                                                                                                                          

\begin{section}{Vanquishing vacuum dispersion}\label{varyingtint}

Given the ever tightening experimental constraints on the vacuum dispersion of light, in addition to the numerous theoretical reasons for preserving a constant speed of light, we now seek a mechanism to suppress vacuum dispersion in quantum gravity on all distance scales. In subsections~\ref{DR2} and~\ref{Area2} we detail a simple mechanism capable of removing vacuum dispersion. %Subsections~\ref{correction} and~\ref{gravup} explore further consequences of this proposal and provide non-trivial tests of its validity. 

%%%%%%%%%%%%%%                                                                                                                                                                           
\begin{subsection}{Removing vacuum dispersion in CDT}\label{DR2}

  In subsection~\ref{dimred} (see also Ref.~\cite{Coumbe:2015zqa}) we showed that a reduction of the spectral dimension, and the associated scale dependent speed of light $c_{m}=\Gamma_{+}(r)$, in CDT can be attributed to a path length that varies as a function of the distance scale $r$ with which one probes spacetime. Let $l$ denote the path length of the diffusing particle as $r \rightarrow \infty$ and $l'$ the path length when probed at a finite scale $r$, then

\begin{equation}\label{ratioL}
\frac{l'}{l}= \sqrt{1+ \frac{A l_{p}^{2}}{r^2}} \equiv \Gamma_{+}(r).
\end{equation}

As an illustrative example we now consider a light-clock that ticks each time the massless diffusing particle traverses the distance between its two parallel mirrors.\interfootnotelinepenalty=10000 \footnote{\scriptsize The mass of the mirrors is assumed to be sufficiently large such that the photons have a negligible affect on the mirror. See~\cite{Ng:2004bt} for details on the practicalities of measuring a spacetime distance using clocks and mirrors.} The particle's path length will increase in response to being probed at a decreasing scale $r$ according to the function $\Gamma_{+}(r)$ defined by Eq.~(\ref{ratioL}). Therefore, if we are to maintain a scale invariant speed of light despite such an increasing path length then the light-clock must compensate by ticking slower according to the same factor $\Gamma_{+}(r)$---time dilates as a function of the relative scale with which we probe the system~\cite{Coumbe:2015bka}.

Defining $t$ as the time it takes a photon to traverse the distance between the mirrors along a geodesic, and $t'$ as the time it takes when probed at a scale $r$, we have

\begin{equation}\label{ratioT}
\frac{t'}{t}= \sqrt{1+ \frac{A l_{p}^{2}}{r^2}} \equiv \Gamma_{+}(r).
\end{equation}

\noindent Thus, in order to maintain a scale invariant speed of light and spectral dimension time must dilate as a function of resolution $r$ according to Eq.~(\ref{ratioT}). We therefore propose that if CDT could be reformulated to include such a scale dependent time dilation in its construction then the observed dimensional reduction and associated superluminality would be removed. The mechanism of scale dependent time dilation was first proposed in Ref.~\cite{Coumbe:2015zqa} (also see Ref.~\cite{Coumbe:2015bka}).

%%%%%%%%%%%%%%%%%%%

We now briefly discuss how this scheme could be implemented in CDT. CDT has a fixed foliation of space-like hypersurfaces that may define a global proper time coordinate, or may just amount to a choice of gauge as argued by Markopoulou and Smolin in $(1+1)$-dimensional CDT~\cite{Markopoulou:2004jz}. For each triangulation there exists a fixed time-like and space-like edge length, and so given the fixed foliation in CDT this means that the time lapse is also fixed. However, if CDT is diffeomorphism invariant the lapses must be able to vary without changing physically observable quantities. One way to implement a scale dependent time interval in CDT may then be to allow the lapse to vary as a function of $\sigma$ according to the proposed scale dependent time dilation. An alternative approach, which may be easier to numerically implement, would be to allow the probability that a diffusing particle transitions to an adjacent time slice to vary as a function of $\sigma$ according to Eq.~(\ref{GammaFuncSig}). It would then be interesting to see if the resulting Monte Carlo simulations yield a scale invariant spectral dimension. %We postpone the implementation of this scheme in CDT to future work. 

%CDT assumes a fixed foliation of space-like hypersurfaces that may be interpreted as defining a global proper time coordinate. One way to implement a scale dependent time coordinate in CDT is to allow the probability that a diffusing particle transitions to an adjacent time slice to vary as a function of its geodesic distance from a randomly chosen origin in accordance with the proposed scale dependent time dilation. It would then be interesting to see if the resulting Monte Carlo simulations yield a scale invariant spectral dimension, or not.

%%%%%%%%%%%%%%%%%%%

In analogy with Eq.~(\ref{ratioT}), we point out that the energy dependent speed of light implied by a reduction of the spectral dimension in CDT becomes energy independent if time also dilates according to

\begin{equation}
\frac{t'}{t}=\sqrt{1+\frac{A B^{2}}{4\pi^{2}}\frac{E^{2}}{E_{P}^{2}}} \equiv \Gamma_{+}(E),
\end{equation}

\noindent where $E$ is the energy required to probe spacetime at a scale $r=(2\pi)/ (B E)$. For $E \ll E_{P}$ we recover the expected result $t'=t$. However, as we increase the energy such that $E  \approx E_{P}$ the time dilation factor $\Gamma_{+}(E)$ begins to significantly deviate from unity, which may modify dynamics at the Planck scale.

\end{subsection}

%%%%%%%%%%%%%%%%%%%%%%%%%%%%%%%%%%%%%%%%%%%%%%%%%%%%%%%%%%%%%%%%%%%%%%%%%%%%%%%%%%%%%%%%%%%%%
%%%%%%%%%%%%%%%%%%%%%%%%%%%%%%%%%%%%%%%%%%%%%%%%%%%%%%%%%%%%%%%%%%%%%%%%%%%%%%%%%%%%%%%%%%%%%

\begin{subsection}{Removing vacuum dispersion given a discrete area spectrum}\label{Area2}

In subsection~\ref{area} we studied how an equidistant area spectrum constrains the propagation of light, finding a scale dependent propagation speed consistent with that observed in the CDT approach to quantum gravity. Crucially, this finding relied on the assumption that time is scale invariant. Hereafter, we remove this assumption and compute how time would have to vary as a function of distance scale in order to be consistent with a scale invariant speed of light.

Recall that in subsection~\ref{area} we considered a spherical wavefront expanding from a point $P$ with an area spectrum given by

\begin{equation}
A_{n}=\frac{4\pi}{B}n l_{P}^{2}.
\label{tcondition1}
\end{equation}

\noindent At a geodesic distance $r_{n}$ from $P$ the wavefront then defines a sphere of area

\begin{equation}
4 \pi r_{n}^{2}=4 \pi c^{2}t_{n}^{2}=\frac{4\pi}{B}n l_{P}^{2},
\label{tcondition1}
\end{equation}

\noindent and at the larger geodesic distance $r_{n+1}$ an area

\begin{equation}
4 \pi r_{n+1}^{2}=4 \pi c^{2}t_{n+1}^{2}=\frac{4\pi}{B} \left(n+1\right)l_{P}^{2},
\label{tcondition2}
\end{equation}

\noindent where $t_{n}$ and $t_{n+1}$ are the time it takes the spherical wavefront to reach a geodesic distance $r_{n}$ and $r_{n+1}$ from the point $P$, respectively (see Fig.~\ref{LightWave}). Equations~(\ref{tcondition1}) and~(\ref{tcondition2}) then give

\begin{equation}
\frac{t_{n+1}}{t_{n}}=\sqrt{1+\frac{1}{B}\frac{l_{P}^{2}}{r_{n}^{2}}} \equiv \Gamma_{+}\left(r_{n}\right).
\label{condition1}
\end{equation}

\noindent The discretisation of area therefore demands that both distance and time be discrete variables, as implied by Eqs.~(\ref{Gamma2}) and~(\ref{condition1}), respectively. However, in the large distance, zero energy, limit they asymptotically approach continuous variables.

Using Eq.~(\ref{condition1}) and defining $\delta t_{n} \equiv t_{n}-t_{n-1}$ and $\delta t_{n+1} \equiv t_{n+1}-t_{n}$ we obtain

\begin{equation}
\frac{\delta t_{n}}{\delta t_{n+1}} = \frac{ 1-\frac{1}{\Gamma_{-}\left(r_{n}\right) }}   {\Gamma_{+}\left(r_{n}\right) -1}.
\label{TfiniteDilation}
\end{equation}

\noindent The scale dependent time dilation of Eq.~(\ref{TfiniteDilation}) then precisely counteracts the scale dependent geodesic distance of Eq.~(\ref{RfiniteDilation}). Hence, if time dilates as a function of distance scale according to Eq.~(\ref{TfiniteDilation}) then we recover a scale invariant speed of light

\begin{equation}
\frac{c\left(r_{n}\right)}{c\left(r_{n+1}\right)}=\frac{\delta r_{n}}{\delta t_{n}}\frac{\delta t_{n+1}}{\delta r_{n+1}}= \frac{ 1-\frac{1}{\Gamma_{-}\left(r_{n}\right) }}{\Gamma_{+}\left(r_{n}\right) -1}   \frac{\Gamma_{+}\left(r_{n}\right) -1}{1-\frac{1}{\Gamma_{-}\left(r_{n}\right) }}=1.
\label{vel2}
\end{equation}

\noindent Similarly, scale dependent time dilation as an infinitesimal ratio 

\begin{equation}
\frac{dt_{n}}{dt_{n+1}}=\Gamma_{+}\left(r_{n}\right)
\label{Ttransform}
\end{equation}

\noindent renders the propagation speed of Eq.~(\ref{diffr}) scale invariant, namely

\begin{equation}
\frac{c\left(r_{n}\right)}{c\left(r_{n+1}\right)}=\frac{dr_{n}}{dt_{n}}\frac{dt_{n+1}}{dr_{n+1}}=\frac{\Gamma_{+}\left(r_{n}\right)}{\Gamma_{+}\left(r_{n}\right)}=1.
\label{vel2}
\end{equation}

%\noindent where we have made use of Eq.~(\ref{diffr}).

\noindent Thus, time must dilate as a function of scale $r_{n}$ relative to the Planck scale $l_{P}$ according to Eq.~(\ref{TfiniteDilation}) (or in the infinitesimal approximation according Eq.~(\ref{Ttransform})) if we are to reconcile the discretisation of area with an invariant speed of light.

\end{subsection}
\end{section}

\begin{section}{Tests of proposed scaling relation}

\begin{subsection}{Comparison with leading quantum correction to the gravitational potential}\label{correction}

As is well-known, gravity as a perturbative quantum field theory is nonrenormalizable. However, general relativity has been successfully formulated as an effective quantum field theory that remains accurate close to its cut-off at the Planck scale, suggesting gravity defines the best behaved effective quantum field theory in nature~\cite{Donoghue:1995cz}. By treating general relativity as an effective quantum field theory, exact leading order quantum corrections to the gravitational potential have been explicitly calculated~\cite{Donoghue:1993eb,Donoghue:1994dn}. These quantum corrections constitute one of the few exact results we know of that a full theory of quantum gravity must reproduce in the appropriate limit. We compare the leading quantum correction of the gravitational potential with what one obtains by applying the scale transformation $r \rightarrow r \Gamma_{+}(r)$ to the classical potential, the purpose of which is to provide a non-trivial test of this transformation. We also predict the next-to-leading order quantum correction to the potential, with the hope that this prediction can be compared with future perturbative calculations.  

The work of Ref.~\cite{Donoghue:1993eb} finds the gravitational potential $V(r)$ between two masses $m_{1}$ and $m_{2}$ including leading quantum corrections in powers of energy, or inverse distance, to be

\begin{equation}\label{loqc}
V(r)=-\frac{G m_{1}m_{2}}{r}\left[1- \frac{G\left(m_{1} +m_{2}\right)}{rc^{2}} - \frac{127}{30\pi^{2}} \frac{G \hbar}{r^{2}c^{3}}  \right].
\end{equation}

\noindent This leading order correction is entirely independent of whatever high energy form quantum gravity takes, and thus constitutes a true prediction of quantum general relativity~\cite{Donoghue:1993eb}. The first correction of order $G/(rc^{2})$ is not a quantum correction as it does not contain any powers of $\hbar$ and can be determined from purely classical considerations, for example by expanding the time component of the Schwarzschild metric~\cite{Donoghue:1993eb}.\interfootnotelinepenalty=10000 \footnote{\scriptsize In this and the following subsection we do not set $\hbar=c=1$.}  

We now apply the scale transformation $r \rightarrow r \Gamma_{+}(r) \equiv \tilde{r}$ to the classical potential 

\begin{equation}
V(r)=-\frac{G m_{1}m_{2}}{r}\left[1- \frac{G\left(m_{1} +m_{2}\right)}{rc^{2}} \right],
\end{equation}

\noindent obtaining

\begin{equation}\label{exactCorrections}
V(\tilde{r})=-\frac{G m_{1}m_{2}}{r \Gamma_{+}(r)}\left[1- \frac{G\left(m_{1} +m_{2}\right)}{r\Gamma_{+}(r)c^{2}} \right].
\end{equation}

\noindent Performing a series expansion of $\Gamma_{+}(r)=\sqrt{1+Bl_{P}^{2}/(\pi r^{2})}$ in powers of $l_{P}$ we find

%\begin{equation}\label{QGcorrection1}
\begin{align}\label{QGcorrection1}
V(\tilde{r}) \approx& -\frac{G m_{1}m_{2}}{r} \bigg[ 1- \frac{G\left(m_{1} +m_{2}\right)}{rc^{2}} - \frac{B}{2\pi} \frac{G \hbar}{r^{2}c^{3}} \\ 
             &+ \frac{B}{\pi} \frac{G^{2} \hbar \left(m_{1} + m_{2}\right)}{r^{3}c^{5}} + \mathcal{O}\left( \hbar^{2}\right) \bigg], \nonumber
%V(\tilde{r}) &=  -\frac{G m_{1}m_{2}}{r}\left[1- \frac{G\left(m_{1} +m_{2}\right)}{rc^{2}} - \frac{1}{2B} \frac{G \hbar}{r^{2}c^{3}}  + \\
%&= \frac{G^{2} \hbar \left(m_{1} + m_{2}\right)}{Br^{3}c^{5}} + \mathcal{O}\left( \hbar^{2}\right) \right].
\end{align} 
%\end{equation}

%V(\tilde{r})=-\frac{G m_{1}m_{2}}{r}\left[1- \frac{G\left(m_{1} +m_{2}\right)}{rc^{2}} - \frac{1}{2B} \frac{G \hbar}{r^{2}c^{3}}  + \frac{G^{2} \hbar \left(m_{1} + m_{2}\right)}{Br^{3}c^{5}} + \mathcal{O}\left( \hbar^{2}\right) \right].
%\end{equation}

\noindent where we have used $l_{P}^{2}=\hbar G/c^{3}$. Equation~(\ref{QGcorrection1}) therefore correctly matches the form of the leading order quantum correction to the gravitational potential. Note that we have also included the leading correction to the term of order $G/(rc^{2})$. 

As a prediction, we now compute the next-to-leading order term, that is the term quadratic in $\hbar$, finding 

\begin{equation}\label{nlo0}
\mathcal{O}\left( \hbar^{2}\right)=-\frac{B^{2}}{\pi^{2}}\frac{G^{3}\hbar^{2}(m_{1}+m_{2})}{B^{2}c^{8}r^{5}} + \frac{3B^{2}}{8\pi^{2}}\frac{G^{2}\hbar^{2}}{c^{6}r^{4}}.
\end{equation}

\noindent Comparing numerical coefficients linear in $\hbar$ sets $B=127 / (15\pi)$, which gives the next-to-leading order term as

\begin{equation}\label{nlo}
\mathcal{O}\left( \hbar^{2}\right)=-\frac{16129}{225 \pi^{4}}\frac{G^{3}\hbar^{2}(m_{1}+m_{2})}{c^{8}r^{5}}+ \frac{16129}{600 \pi^{4}}\frac{G^{2}\hbar^{2}}{c^{6}r^{4}}.
\end{equation}

\noindent Of course, the exact numerical coefficients in Eq.~(\ref{nlo}) depend on the coefficient of the leading quantum correction in Eq.~(\ref{loqc}) being correct, which has been questioned by Ref.~\cite{Khriplovich:2002bt}. Regardless, whatever the correct prefactor of the leading quantum correction, one can equate it with $B/ (2\pi)$ and compute the explicit quantum correction to any order by making a series expansion of $\Gamma_{+}(r)$ in Eq.~(\ref{exactCorrections}). 

\end{subsection}

%%%%%%%%%%%%%%%%%%%%%%%%%%%%%%

\begin{subsection}{Comparison with the generalised uncertainty principle}\label{gravup}

%The uncertainty principle is now fundamental to our understanding of physical phenemona. 
Heisenberg first derived a version of the position-momentum uncertainty principle by considering how to measure the position of an electron using a microscope. The precision with which one can determine the position of an electron $\Delta x$ is limited by the wavelength $\lambda$ of the electromagnetic wave one uses to make the measurement, $\Delta x\approx \lambda$. Due to the quantization of the electromagnetic field into photons with discrete momenta $p=\hbar / \lambda$, a photon scattering from an electron must impart a non-zero component of its momentum to the electron, thereby introducing a non-zero uncertainty $\Delta p\approx \hbar / \Delta x$. Thus, the position-momentum uncertainty principle is an inescapable consequence of making measurements using discrete momentum carrying particles. The standard position-momentum uncertainty principle is given by

\begin{equation}\label{up}
\Delta p \geq \frac{\hbar}{2 \Delta x}.
\end{equation}

However, Eq. (\ref{up}) is almost certainly incomplete, as it does not account for gravitational interactions induced by the act of observation~\cite{Mead:1964zz,Maggiore:1993rv,Garay:1994en,Adler:1999bu}. Since any scattering particle must have a non-zero momentum it must also have a non-zero energy. The gravitational field of the scattering particle must then cause a non-zero acceleration of the electron, thus perturbing its original position. This gravitational contribution to uncertainty seems an inevitable result of combining general relativity and quantum mechanics, and is consequently considered a likely feature of quantum gravity.  

This so-called generalised uncertainty principle (GUP) was originally derived within the context of superstring theory \cite{Veneziano:1986zf}, and has since been obtained independent of any particular approach to quantum gravity. Numerous model-independent \emph{gedanken} experiments~\cite{Maggiore:1993rv,Scardigli:1999jh} find a functional form for the GUP that is identical to the results obtained within the framework of string theory~\cite{Veneziano:1986zf,Amati:1988tn,Konishi:1989wk}. Moreover, dimensional analysis and explicit calculations using Newtonian gravity and general relativity find the same modification of the standard uncertainty principle~\cite{Adler:1999bu}. The fact that such a large number of independent derivations converge on the same form of the GUP, with many arguments assuming nothing more than general relativity and quantum mechanics, strongly suggests the GUP is a robust prediction of quantum gravity. The position-momentum GUP has been found to take the general form

\begin{equation}\label{gup}
\Delta x \Delta p \approx \frac{\hbar}{2} + \alpha \frac{l_{P}^{2}\Delta p^{2}}{\hbar},
\end{equation}

\noindent where $\alpha$ is an unknown constant of proportionality~\cite{Adler:1999bu}. An important feature of the GUP is that it has a minimum position uncertainty which can be interpreted as a minimum distance scale~\cite{Adler:1999bu}. 

We now compare the GUP of Eq.~(\ref{gup}) with the result of transforming position uncertainty in the standard Heisenberg relation of Eq.~(\ref{up}) according to $\Delta x \rightarrow \Delta x \Gamma_{+}(\Delta x) \equiv \Delta \tilde{x}$, hence

\begin{equation}\label{gup1}
\Delta \tilde{p} \geq \frac{\hbar}{2 \Delta \tilde{x}} \geq \frac{\hbar}{2 \Delta x \Gamma_{+}(\Delta x)}.
\end{equation}

\noindent Performing a series expansion of $\Gamma_{+}(\Delta x)$ in powers of $l_{P}$ and using the approximation $\Delta x=\hbar /(2 \Delta p)$ (which is valid for $\Delta x \gg l_{P}$ or if $\Delta p$ transforms according to $\Delta p \rightarrow \Delta p / \Gamma_{+}(\Delta x) \equiv \Delta \tilde{p}$) we obtain

\begin{equation}\label{gup2}
\Delta x \Delta \tilde{p} \geq \frac{\hbar}{2\Gamma_{+}(\Delta x)} \approx \frac{\hbar}{2} - B\frac{l_{P}^{2}\Delta p^{2}}{\pi \hbar} + \mathcal{O}\left(l_{P}^{4}\right).
\end{equation}

\noindent Equations~(\ref{gup}) and~(\ref{gup2}) therefore have the same functional form.   

One of the principle reasons to expect vacuum dispersion in quantum gravity comes from the standard Heisenberg uncertainty relation, which says that the smaller the region of spacetime under consideration the greater the magnitude of allowed energy fluctuations. Viewed at sufficiently small distance scales energy fluctuations can then become large enough to induce significant fluctuations of the spacetime metric $\delta g_{\mu\nu}$. A low energy photon for which $\lambda \gg \delta g_{\mu\nu}$ will not resolve any such metric fluctuations, and its path length $l$ will remain unperturbed. However, a higher energy photon for which $\lambda \approx \delta g_{\mu\nu}$ will begin to resolve the vacuum fluctuations, inducing a non-zero perturbation of the path length $l'=l+\delta l$. 

%The ratio of the path length for high energy photons $l'$ and low energy photons $l$ is then                                                                                                                                                             

%\begin{equation}\label{deltal0}                                                                                                                                                           % 
%\frac{l'}{l}=1+\frac{\delta l}{l}.                                                                                                                                                         
%\end{equation}                                                                                                                                                                             

Vacuum fluctuations are thought to only become significant on extremely small distance scales ($l_{P}\sim 10^{-35}m$), however, it is expected that their accumulative effect over cosmological scales should be observable. Although approaches to quantum gravity differ on the exact size of the path length fluctuations $\delta l$ it is possible to parametrise how they are expected to accumulate over a distance $l$ in a model independent way. This is done by defining a free parameter $\alpha$, such that                                                          

\begin{equation}\label{deltal}                                                                                                                                                             
\delta g_{\mu\nu} \approx \delta l \simeq l^{1-\alpha}l_{p}^{\alpha}, 
\end{equation}                                                                                                                                                                             

\noindent where $\alpha$ can in principle be determined by experiment and compared with predictions from various approaches to quantum gravity~\cite{Perlman:2014cwa}. For example, the so-called random walk model predicts $\alpha=1/2$ \cite{Vasileiou:2015wja,Diosi:1989hy} and a restricted version of holography predicts $\alpha=2/3$ \cite{Perlman:2014cwa}. Reference \cite{Perlman:2014cwa} analyses data from the \emph{Chandra} and \emph{Fermi} space telescopes in addition to the ground based \emph{Cherenkov} telescopes, finding that values of $\alpha \leq 0.72$ are experimentally excluded.     

Rewriting Eq.~(\ref{gup2}) as 

\begin{equation}\label{gup3}
\Delta \tilde{p} \geq \frac{\hbar}{2\Delta \tilde{x}},
\end{equation}

\noindent we see that $\Delta \tilde{p}$ is maximum when $\Delta \tilde{x}$ is minimum. Since $\Delta \tilde{x}$ has a minimum of $l_{P} \sqrt{B/ \pi}$ the maximum momentum uncertainty is 

\begin{equation}\label{gup3}
\Delta \tilde{p}_{max}=\frac{1}{2}\sqrt{\frac{\pi}{B}}\frac{E_{P}}{c}.
\end{equation}

\noindent Therefore, Eq.~(\ref{gup3}) tells us that vacuum fluctuations with an energy greater than $\approx$$E_{P}$ are forbidden, meaning that deviations from a smooth manifold will be negligible, if not vanishing. 

We can now intuitively summarise how a scale dependent time interval allows a photon’s speed to remain independent of its energy via the uncertainty principle. Since shorter wavelength light probes spacetime on smaller distance scales it should encounter proportionately larger energy fluctuations. However, since shorter wavelength light also experiences a larger time dilation factor it will also observe the dynamical fluctuations to be proportionately suppressed. The result --- the observed magnitude of vacuum fluctuations is wavelength independent, allowing a photon's speed to always remain independent of its energy.

%\begin{equation}\label{gup3}
%\Delta \tilde{x} \Delta p \geq \frac{\hbar}{2} - \frac{l_{P}^{2}\Delta p^{2}}{\hbar B} + \mathcal{O}\left(l_{P}^{4}\right).
%\end{equation}

\end{subsection}

\end{section}

%%%%%%%%%%%%%%%%%%%%%%%%%%%%%%%%%%%%%%%

%%%%%%%%%%%%%%%%%%%%%%%%%%%%%%%%%%%%%%%%%%%%%                                                                                                                                              

\begin{section}{Conclusions}

In this work we have improved on the evidence first presented in Ref.~\cite{Coumbe:2015zqa} for a scale dependent speed of light in CDT quantum gravity by performing numerical measurements using two different lattice volumes and by averaging over a greater number of diffusion processes. We find evidence to support the claim that the effective velocity of diffusing particles in CDT becomes superluminal on small distance scales, and characterise the functional form of vacuum dispersion in CDT via fits to the numerical data. We show that the observed scale dependent speed of light in CDT can be accounted for by a scale dependent transformation of geodesic distance, whose specific functional form implies a discrete equidistant area spectrum. We make two non-trivial tests of the proposed scale transformation: a comparison with the leading-order quantum correction to the gravitational potential~\cite{Donoghue:1993eb} and a comparison with the generalised uncertainty principle~\cite{Adler:1999bu}. In both comparisons, we obtain excellent agreement, and in the case of the gravitational potential we even predicted the next-to-leading order quantum correction. We point out that the same functional form of scale dependent geodesic distance has been independently derived in Refs.~\cite{padmanabhan1985physical,Kothawala:2013maa,Padmanabhan:2015vma} within a different context.   

We wish to briefly clarify the modification of geodesic distance proposed in this work and discuss its possible limitations. Primarily motivated by numerical calculations of the spectral dimension we propose that quantum gravitational vacuum fluctuations modify the classical notion of an exact geodesic distance $r$ between two spacetime points $x_{1}$ and $x_{2}$ in all inertial frames of reference to 

\begin{equation}\label{geo1}
\tilde{r}^{2}(x_{1},x_{2}) = r^{2}(x_{1},x_{2}) \left[ 1 + \frac{\zeta l_{P}^{2}}{\lambda^{2}}\right ],
\end{equation}

\noindent where $\zeta$ is a constant of proportionality, $l_{P}$ is the Planck length and $\lambda$ is the distance scale with which one probes spacetime. The spectral dimension, as defined in this work, is strictly only valid for infinitely flat Euclidean space, although in principle one can still use this definition to compute the fractal dimension of a curved space by factoring in the appropriate corrections (see Refs.\cite{Ambjorn:2005db,Coumbe:2014noa}). Therefore, it is important to note that the modification of geodesic distance proposed in this work may only be exact for an infinite flat spacetime or for a local region of a curved manifold, where the various definitions of geodesic distance explored in this work should also coincide.

%By considering the propagation of light constrained by the discretisation of area we also obtain an identical functional form for the scale dependent speed of light, which may suggest a connection between the appearance of dimensional reduction and the discretisation of area. The energy dependent speed of light in both cases agrees with our expectations for vacuum dispersion based on a series expansion at small energies and using effective field theory~\cite{Vasileiou:2013vra}.  

%We show that the appearance of dimensional reduction and the discretisation of area both imply a specific scale dependent geodesic interval, and explore the conservative possibility that such a scaling may result from simple general relativistic considerations. 

%We show that the observed scale dependent speed of light in CDT can be accounted for by a scale dependent transformation of geodesic distance, whose specific functional form implies the discretisation of spherical surface area.

%As a non-trivial test we compare the known leading quantum correction to the gravitational potential~\cite{Donoghue:1993eb} with what one obtains by applying a scale dependent geodesic interval to the classical potential. A comparison is also made with the so-called generalized uncertainty principle (GUP)~\cite{Adler:1999bu}. In both cases the comparison yields identical functional forms. 

Assuming such a scale dependent geodesic interval, a scale dependent time interval becomes essential to maintaining an invariant speed of light.\interfootnotelinepenalty=10000 \footnote{\scriptsize Ref.~\cite{Polchinski:2011za} also briefly highlights that by discretising space and time with the same scale one can prevent low-energy Lorentz violation.} We find that a scale dependent time interval is all that is needed to eliminate the vacuum dispersion associated with the specific case of dimensional reduction, and detail how it may reconcile more general predictions of vacuum dispersion in quantum gravity with the growing body of contrary experimental evidence. 

%We explore the possibility that gravitational time dilation may naturally account for the necessary scale dependent time interval. 

%We briefly discuss how this scenario permits an observer independent minimal length scale and an invariant speed of light, as per the aims of doubly special relativity~\cite{AmelinoCamelia:2002wr}.

%One of the main reasons to expect vacuum dispersion in quantum gravity is discussed in some detail, namely the putative existence of a minimal length scale. In particular, we discuss a possible modification of the Lorentz transformations that are consistent with an observer independent speed of light and a minimal length scale, in addition to being consistent with measurements of dimensional reduction and the discretisation of area. 

In physics the arbiter of truth is experiment, and every experiment ever performed has proved consistent with precisely zero vacuum dispersion, now even beyond the Planck scale~\cite{Vasileiou:2013vra,FermiGBMLAT1,Nemiroff:2011fk}. Recent experimental constraints are increasingly motivating the need to adapt or eliminate approaches to quantum gravity that predict the vacuum dispersion of light. It is, therefore, important to investigate mechanisms capable of accounting for the absence of vacuum dispersion at the Planck scale. This work demonstrates how quantum gravity might be reconciled with precisely zero vacuum dispersion by allowing time to dilate as a function of the distance scale with which one probes spacetime. The prediction of an energy independent speed of light is falsifiable with current experimental sensitivities using a variety of different techniques~\cite{Vasileiou:2013vra,FermiGBMLAT1,Nemiroff:2011fk,Bi:2008yx}, and the predicted next-to-leading order quantum correction to the gravitational potential of Eq.~(\ref{nlo}) may be falsifiable with higher-order perturbative calculations. 

\end{section}

%%%%%%%%%%%%%%%%%%%%%%%%%%%%%%%%%%%%%%%

\section*{Acknowledgements}

I wish to thank Jan Ambjorn and Jerzy Jurkiewicz for discussions and comments on the manuscript, and acknowledge support from the ERC-Advance grant 291092, ``Exploring the Quantum Universe'' (EQU).

\bibliographystyle{unsrt}
\bibliography{Master}

%%%%%%%%%%%%%%%%%%%%%%%%%%%%%%%%%%%%%%% 

\end{multicols}

\end{document}